\documentclass[11pt]{article}
\usepackage{graphicx}

\title{Finite temperature dynamic structure function of the free Fermi
gas} 

\author{F.Mazzanti$^1$ and A.Polls$^2$ \\  \\
          $^1$ {\small Departament d'Electr\`onica,
                       Enginyeria i Arquitectura La Salle,} \\
               {\small Pg. Bonanova 8, Universitat Ramon Llull,} \\
               {\small E-08022 Barcelona, Spain} \\ 
          $^2$ {\small Departament d'Estructura i Constituents
                       de la Mat\`eria,} \\
               {\small Diagonal 645, Universitat de Barcelona,} \\
               {\small E-08028 Barcelona, Spain} }

\begin{document}

\maketitle

\begin{abstract}
A detailed discussion of the coherent and incoherent dynamic structure
function of the free Fermi gas at finite temperature is
presented. Their behavior and evolution with the momentum transfer and
the temperature is analyzed, while particular attention is paid to the
way in which their relative contribution changes with respect to the
$\tilde{T}=0$ case. The influence of thermal effects on the lowest
order sum rules of the coherent and incoherent responses is also
discussed. Finally, the scaling properties of the responses at high
momentum transfer as a function of the temperature are also analyzed. 
\end{abstract}

\maketitle

\pagebreak

Neutron scattering in quantum liquids at low and high momentum
transfer $q$ has provided much relevant information about the role of
collective excitations, the way in which single--particle properties
are affected by correlations and the nature of quantum effects in
general.  Theoretical models devised to explain the observed
scattering data have also been developped, and the degree of
sophistication achieved is such that nowadays precise understanding of
many features concerning the ground state of systems like $^4$He,
$^3$He or $^3$He--$^4$He mixtures has been gained~\cite{gly1}.
Although Path Integral Monte Carlo has supplied unique information
about finite temperature effects in these systems, no particular
formalisms has been capable to describe the dynamic structure function
$S(q,\omega)$ at $T>0$~\cite{ceper1}. In a previous work we presented
a detailed description of the $T=0$ dynamic structure function of the
free Fermi gas, its coherent and incoherent parts and their evolution
with the momentum transfer~\cite{mazzpo1}. In the present letter we
extend this discussion to finite temperature, focussing particularly
on the influence of thermal effects on the different contributions to
the response and on scaling laws.

At finite temperature, the dynamic structure function of a quantum
system is proportional to the probability of coupling different
states that are compatible with a momentum and energy transfer $q$ and
$\omega$ when a density fluctuation moderated by the operator
$\rho_q=\sum_{j=1}^N e^{i{\bf q}\cdot{\bf r}_j}$ takes place
\begin{equation}
S(q,\omega) = \sum_{\{n,m\}} \frac{1}{\mathcal{Z}} 
e^{-\beta (E_n-\mu N)} \frac{1}{N}
\left| \left\langle m \mid \rho_q \mid n \right\rangle \right|^2
\delta\left( E_m - E_n - \omega \right) \ ,
\label{sqw-b1}
\end{equation}
${\mathcal{Z}}=\sum_{\{n\}}e^{-\beta (E_n-\mu N)}$ being the partition
function of the system in the Grand Canonical ensamble, $\mu$ the
chemical potential and $\beta=1/k_BT$ the inverse of the 
temperature~\cite{pin1}.

The dynamic structure function is also the Fourier transform of the
density--density correlation factor $S(q,t)$, which can in turn be
formally separated in its coherent and incoherent density
responses. At finite temperature these two functions are defined as
\begin{eqnarray}
S_{inc}(q,t) \!\!\! & = & \!\!\! 
\sum_{\{n\}} \frac{1}{\mathcal{Z}}
e^{-\beta (E_n-\mu N)} \frac{1}{N} \sum_{j=1}^N \left\langle n \mid
e^{-i{\bf q}\cdot{\bf r}_j} e^{iHt} e^{i{\bf q}\cdot{\bf r}_j}
e^{-iHt} \mid n \right\rangle 
\label{sqw-b2} \\ [2mm]
S_{coh}(q,t) \!\!\! & = & \!\!\! 
\sum_{\{n\}} \frac{1}{\mathcal{Z}}
e^{-\beta (E_n-\mu N)} \frac{1}{N} \sum_{i\neq j}^N \left\langle n \mid
e^{-i{\bf q}\cdot{\bf r}_i} e^{iHt} e^{i{\bf q}\cdot{\bf r}_j}
e^{-iHt} \mid n \right\rangle \ ,
\label{sqw-b3}
\end{eqnarray}
where $H$ and ${\bf r}_j$ are the Hamiltonian and the position operator
of particle $j$, respectively. 
Notice that in this definition of $S_{coh}(q,\omega;T)$ \cite{mazzpo1,
momdist,gersch}
,only those
terms with $i$ strictly different from $j$ contribute to the coherent
response, in contrast to the other commonly used definition where 
$S_{coh}(q,\omega)$ is taken to be the whole scattering function 
given in Eq.~(\ref{sqw-b1})~\cite{gly1,lovesey}. 
Notice also that in $S_{inc}(q,t)$
correlations among different particles are indirectly accounted by
correlations in the wave functions. In a free classical system where
neither dynamical nor statistical correlations exist, $S_{coh}(q,t)$
is zero and the total response equals its incoherent part. However, in
realistic systems this condition is only asymptotically reached at
high temperature or in the high momentum transfer limit. When
$q\to\infty$, the total response of a fully correlated, infinite and
isotropic system becomes mostly incoherent and conforms to the Impulse
Approximation (IA). In terms of the new set of adimensional variables
that will be used throughout this work $\tilde{q}=q/k_F$,
$\tilde{\omega}=\omega/\epsilon_F$ and $\tilde{T}=T/\epsilon_F$ where
$k_F$ is the $T=0$ Fermi momentum of the free Fermi gas and
$\epsilon_F=k_F^2/2m$, the adimensional IA reads
\begin{equation}
\tilde S_{IA}(\tilde{q},\tilde\omega; \tilde{T}) \equiv
\epsilon_F S_{IA}(q, \omega; T) =
\frac{3}{4\pi} \int d\tilde{\bf k} \,n(\tilde{k})
\,\delta\!\left[ (\tilde{\bf k}+\tilde{\bf q})^2 - \tilde{k}^2 -
\tilde{\omega} \right] = 
\frac{3}{4\tilde q} \int_{\mid \tilde Y\mid}^\infty
\tilde{k}n(\tilde{k})\,d\tilde{k} \ ,
\label{sqw-c1}
\end{equation}
where $\tilde{Y}=(\tilde\omega/\tilde{q}-\tilde{q})/2$ is the West
scaling variable~\cite{west} and $n(\tilde{k})$ is the finite
temperature occupation probability of each single--particle state of
definite momentum. Corrections to the IA are moderated by the
interatomic potential, so in a free system the IA exactly coincides
with $S_{inc}(\tilde{q},\tilde\omega)$ at any value of $\tilde{q}$ and
$\tilde\omega$~\cite{momdist}.

The dynamic structure function is also related to the imaginary part
of the dynamic susceptibility
$\chi(\tilde{q},\tilde\omega)$~\cite{pin1}, which has been calculated
at finite temperature for the free Fermi gas in
ref.~\cite{gly2} and reads
\begin{equation}
\tilde{S}(\tilde{q}, \tilde{\omega}; \tilde{T}) =
\frac{3\tilde{T}}{8\tilde{q}} 
\frac{1}{(1-e^{-\tilde{\omega}/\tilde{T}})}
\ln\left( 
\frac{1+ze^{-\frac{1}{4\tilde{T}}\left(\frac{\tilde\omega}{\tilde{q}}
-\tilde{q}\right)^2}}
{1+ze^{-\frac{1}{4\tilde{T}}\left(\frac{\tilde\omega}{\tilde{q}}
+\tilde{q}\right)^2}} \right) \ ,
\label{sqw-d1}
\end{equation}
where $z=e^{\tilde{\mu}/\tilde{T}}$ is the adimensional fugacity at
temperature $\tilde{T}$ and $\tilde\mu=\mu/\epsilon_F$ is the
adimensional chemical potential. This last quantity can be determined
from the particle normalization condition
\begin{equation}
\frac{2}{3} \tilde{T}^{-3/2} = \int_0^\infty 
\frac{\epsilon^{1/2} d\epsilon}{z^{-1} e^{\epsilon}+1} \ ,
\label{sqw-d2}
\end{equation}
which only depends on the temperature and therefore yields a density
and mass independent chemical potential. Notice that at finite 
$\tilde{T}$, a system can deexcite giving some energy to the probe;
as a consequence the response is also deffined at negative
energies. Actually in thermodynamical equilibrium, the response 
satisfies detailed balance
\begin{equation}
\tilde{S}(\tilde{q}, -\tilde\omega; \tilde{T}) = 
e^{-\tilde\omega/\tilde{T}}
\tilde{S}(\tilde{q}, \tilde\omega; \tilde{T}) \ ,
\label{sqw-d2b}
\end{equation}
as can be straightforwardly checked in Eq.~(\ref{sqw-d1}).

The adimensional incoherent response of a free system can be computed
in the Impulse Approximation
\begin{equation}
\tilde{S}_{inc}(\tilde{q}, \tilde\omega; \tilde{T}) = 
\frac{3\tilde{T}}{8\tilde{q}} \ln \left( 1 + 
ze^{-\frac{1}{4\tilde{T}}\left(\frac{\tilde\omega}{\tilde{q}}
-\tilde{q}\right)^2} \right)
\label{sqw-d3}
\end{equation}
while the coherent response $\tilde{S}_{coh}(\tilde{q}, \tilde\omega;
\tilde{T})$ is the difference between~(\ref{sqw-d1})
and~(\ref{sqw-d3}). Notice that for the free Fermi gas
$\tilde{S}(\tilde{q}, \tilde\omega; \tilde{T})$ can also be obtained
from the well known relation
\begin{equation}
\tilde{S}(\tilde{q}, \tilde\omega; \tilde{T}) =
\frac{3}{4\pi} \int d\tilde{\bf k} \,n(\tilde{k}) \left[ 1 - 
n(\tilde{\bf k}+\tilde{\bf q}) \right] 
\delta\left( (\tilde{\bf k}+\tilde{\bf q})^2 - \tilde{k}^2 -
\tilde\omega \right) 
\label{sqw-d3b} 
\end{equation}
where the finite temperature momentum distribution is
\begin{equation}
n(\tilde{k}) = \frac{1}{z^{-1}e^{\tilde{k}^2/\tilde{T}}+1} \ ,
\label{sqw-e1}
\end{equation}
which yields occupation numbers between 0 and 1. A direct consequence
of the previous deffinitions is that $\tilde{S}_{inc}(\tilde{q},
\tilde\omega; \tilde{T})$ is positive, has no nodes and is symmetric
around the point $\tilde\omega=\tilde{q}^2$ where it shows a peak. In
much the same way, $\tilde{S}(\tilde{q}, \tilde\omega; \tilde{T})$ is
always lower or equal to $\tilde{S}_{inc}(\tilde{q},
\tilde\omega; \tilde{T})$, and therefore $\tilde{S}_{coh}(\tilde{q},
\tilde\omega; \tilde{T})$ is negative and does not change
sign. Furthermore, the coherent response is an even function of the
energy and has its minimum at $\tilde\omega=0$. Finally, 
$\tilde{S}(\tilde{q}, \tilde\omega; \tilde{T})$ and
$\tilde{S}_{inc}(\tilde{q}, \tilde\omega; \tilde{T})$ are 
positive functions satisfying 
$\tilde{S}(\tilde{q}, \tilde\omega; \tilde{T}) \leq 
\tilde{S}_{inc}(\tilde{q}, \tilde\omega; \tilde{T})$, therefore 
the absolute value of 
$\tilde{S}_{coh}(\tilde{q}, \tilde\omega; \tilde{T})$ is always lower
or equal to $\tilde{S}_{inc}(\tilde{q}, \tilde\omega; \tilde{T})$.

\begin{figure}
\begin{center}
\includegraphics[height=16cm]{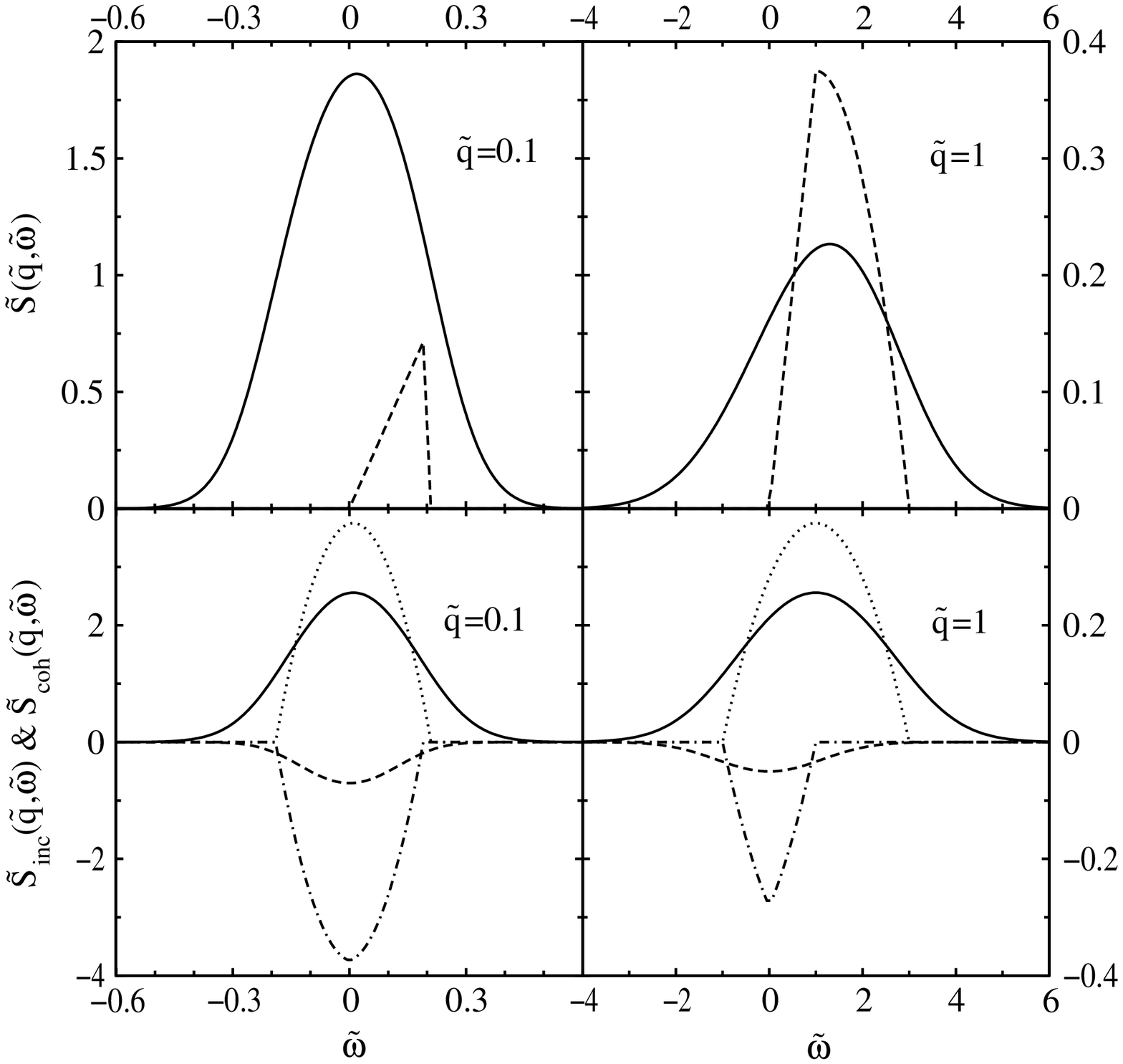}
\caption{Total response and its coherent and incoherent parts
at $\tilde{T}=0$ and $\tilde{T}=1$. Upper plots: total response at
$\tilde{T}=1$ (solid line) and $\tilde{T}=0$ (dashed lines) at
$\tilde{q}=0.1$ and $\tilde{q}=1$. Lower plots: incoherent responses
at $\tilde{T}=1$ (solid line) and at $\tilde{T}=0$ (dotted lines), and
coherent responses at $\tilde{T}=1$ (dashed line) and $\tilde{T}=0$
(dot--dashed line).}
\label{fig-1}
\end{center}
\end{figure}

The $\tilde{q}=0.1$ and $\tilde{q}=1$ total, coherent and incoherent
responses of the free Fermi gas are compared at $\tilde{T}=1$ and
$\tilde{T}=0$ in Fig.~(\ref{fig-1}), where the zero temperature
responses are taken from ref.~\cite{mazzpo1} 
\begin{equation}
\tilde S(\tilde q, \tilde\omega; 0) = 
\left\{ \begin{array}{ll}
		\left. \begin{array}{ll}
			{3\tilde\omega\over 8\tilde q} & 
			\mbox{if\,\,\,$2\tilde{q}-\tilde{q}^2\geq
			     \tilde\omega\geq 0$} \\ [2mm]
			{3\over 8\tilde{q}}\left[1-{1\over 4}\left(
			     {\tilde\omega\over\tilde q}-\tilde q
			     \right)^2\right] & 
			\mbox{if\,\,\,$2\tilde q+\tilde q^2\geq 
			     \tilde\omega\geq 2\tilde q-\tilde q^2$} 
		\end{array} \right\} & \tilde q\leq 2. \\ [8mm]
		\begin{array}{ll}
			{3\over 8\tilde q}\left[1-{1\over 4}\left(
			     {\tilde\omega\over\tilde q}-\tilde q
			     \right)^2\right] & 
			\mbox{if\,\,\,$\tilde q^2+2\tilde q\geq
			     \tilde\omega\geq \tilde q^2-2\tilde q$}
		\end{array} & \mbox{$\tilde q\geq 2$}. \\ [2mm]
		\begin{array}{ll}
			0 \,\,\,\,\,\mbox{otherwise\ ,}
		\end{array} & 
	    \end{array}
   	\right.
\label{sqw-d4}
\end{equation}
and
\begin{equation}
\tilde{S}_{inc}(\tilde{q}, \tilde\omega; 0) = 
\frac{3}{8\tilde{q}} \left[ 1 - \left(
\frac{\tilde\omega}{\tilde{q}} - \tilde{q} \right)^2 \right] \ .
\label{sqw-d5}
\end{equation}

At the momenta considered, both the coherent and the incoherent
responses are finite and contribute to the total dynamic structure
function. As it can be seen from the figure, temperature effects
noticeably quench $\tilde{S}_{inc}(\tilde{q},\tilde\omega;\tilde{T})$
and $\tilde{S}_{coh}(\tilde{q},\tilde\omega;\tilde{T})$ even though
the net effect on the latter is stronger, thus leading to a total
response that is more incoherent at finite temperature than at
$\tilde{T}=0$. This effect turns out to be particularly relevant at
low $\tilde{q}$'s, where the reduction in $\tilde{S}_{coh}(\tilde{q},
\tilde\omega; \tilde{T})$ is large compared to the reduction in
$\tilde{S}_{inc}(\tilde{q}, \tilde\omega; \tilde{T})$ and the total
response becomes strongly enhanced with respect to
$\tilde{S}(\tilde{q},\tilde\omega; 0)$.  The quenching of the
incoherent response may be understood recalling that at finite
temperature single--particle states are occupied according to
Eq.~(\ref{sqw-e1}), and while particle number conservation requires
the second $\tilde{k}$--weighted moment of $n(\tilde{k})$ not to
change with temperature, the way in which the different states above
and below the Fermi level are filled at $\tilde T>0$ is such that the
total contribution of $\tilde{k} n(\tilde{k})$ at fixed $\tilde\omega$
is lower than at $\tilde{T}=0$. Much more significant is the quenching
suffered by the coherent response at low energies, which is due to the
simultaneous action of two effects. On one hand and as commented
above, $\tilde{S}_{coh}(\tilde{q},\tilde{\omega};\tilde{T})$ is always
smaller than $\tilde{S}_{inc}(\tilde q, \tilde\omega;\tilde T)$ in
absolute value, and the latter is at finite temperature lower than at
$\tilde T=0$. On the other, at finite temperature and low momentum
transfer the occupation probability of the states at which particles
transite when the system is given a net momentum $\tilde{\bf q}$ is
not $1$, and hence the effect of Pauli correlations is weakened
compared to the $\tilde{T}=0$ case.

As it can also be seen in the figure, the energy interval covered by
the three responses at finite temperature is larger than the range at
$\tilde{T}=0$. According to eqs.~(\ref{sqw-c1}) and~(\ref{sqw-e1}),
$\tilde{S}_{inc}(\tilde{q}, \tilde{\omega}; \tilde{T})$ contributes at
all energies (both positive and negative) because in the IA the
allowed transitions have energy $\tilde{\omega}=(\tilde{\bf
k}+\tilde{\bf q})^2 -\tilde{k}^2$, and at finite temperature
$n(\tilde{k})$ may extend up to infinity. In contrast, the
$\tilde{T}=0$ occupation is restricted to states with $\tilde{k}\leq
1$ and so the energy range covered by $\tilde{S}_{inc}(\tilde{q},
\tilde{\omega}; 0)$ at fixed $\tilde{q}$ reduces to the interval
$\tilde{\omega}\in\left(\tilde{q}^2-2\tilde{q},
\tilde{q}^2+2\tilde{q}\right)$. Consequently, thermal excitations
promote particles to states above the Fermi level and these contribute
to the large energy tails of the response, which therefore become
enhanced with respect to the $\tilde{T}=0$ case.

It is also important to notice that the incoherent response takes into
account the occupation of the initial states but disregards any
possible constrain on the occupation of the final states at which the
excited particles transite. As symmetry requirements restrict the set
of particle configurations that can be realized in the final state,
the coherent response subtracts from the incoherent one the
contribution of all those transitions that are actually forbidden by
the Pauli exclussion principle, thus leading to a total response in
which correlations in both the initial and the final states are
properly treated. That is the reason why at fixed temperature and
momentum transfer $\tilde{S}_{coh}(\tilde{q}, \tilde{\omega};
\tilde{T})$ is negative and lower or equal to
$\tilde{S}_{inc}(\tilde{q}, \tilde{\omega}; \tilde{T})$ in absolute
value. Finally, as $\tilde{S}_{coh}(\tilde{q}, \tilde{\omega};
\tilde{T})$ subtracts strength from $\tilde{S}_{inc}(\tilde{q},
\tilde{\omega}; \tilde{T})$, and the latter extends over an energy
interval larger than the range covered at $\tilde{T}=0$, the same
happens to $\tilde{S}_{coh}(\tilde{q},\tilde{\omega};\tilde{T})$. In
this sense, the tails of the finite temperature coherent response are
enhanced when compared to 
$\tilde{S}_{coh}(\tilde{q}, \tilde{\omega}; 0)$.

\begin{figure}
\begin{center}
\includegraphics[height=16cm]{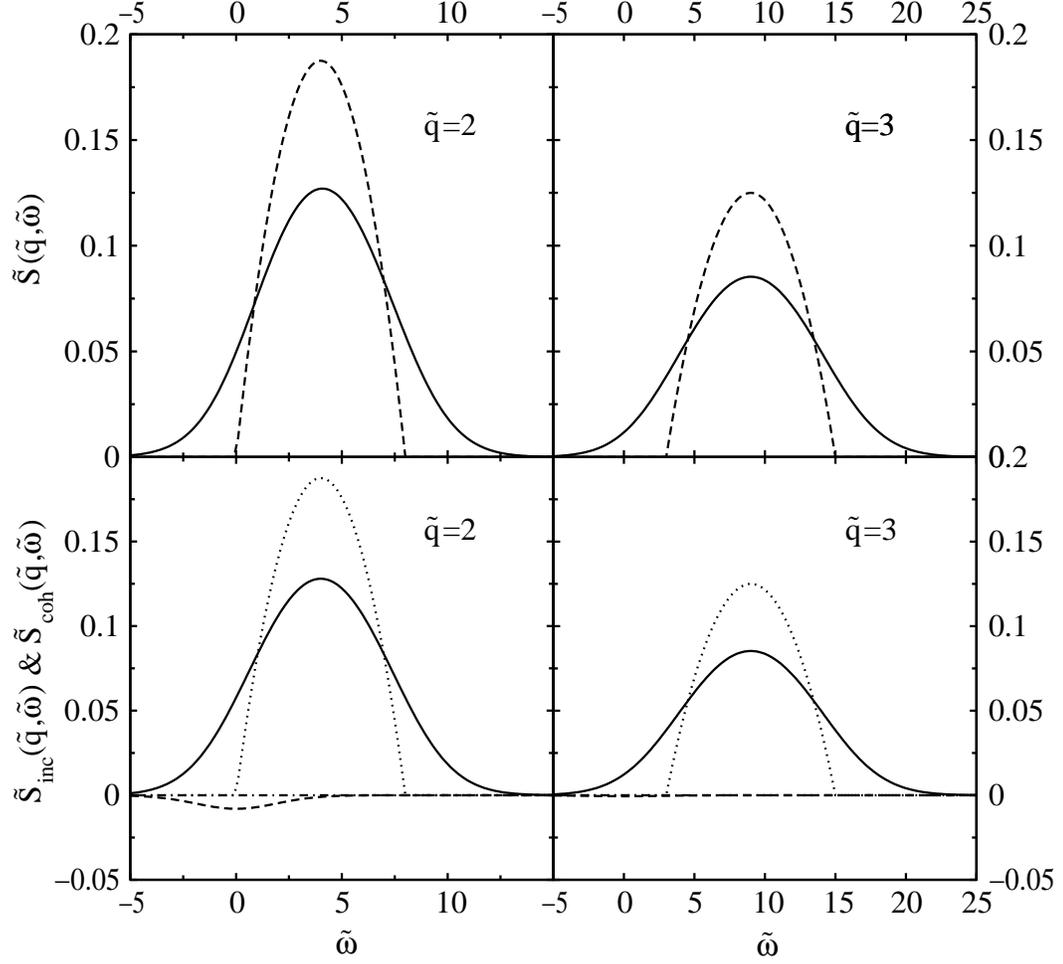}
\caption{Total response and its coherent and incoherent parts
at $\tilde T=0$ and $\tilde T=1$, for $\tilde{q}=2$ and 
$\tilde{q}=3$, with the same notation used in Fig.~(\ref{fig-1}).}
\label{fig-2}
\end{center}
\end{figure}

However, and as it is shown in Fig.~(\ref{fig-2}), a completely
different situation is found when the transferred momentum $\tilde{q}$
is equal or larger than 2. At $\tilde q>2$, $\tilde S_{coh}(\tilde
q,\tilde\omega; 0)$ cancels and the total response becomes entirely
incoherent, while at finite temprature neither $\tilde S_{inc}(\tilde
q,\tilde\omega; \tilde T)$ nor $\tilde S_{coh}(\tilde q,\tilde\omega;
\tilde T)$ vanish and hence $\tilde S(\tilde q, \tilde\omega; \tilde
T)$ has contributions from both functions.  In this sense, therefore,
temperature acts as a source of coherentness, even though the total
response is clearly dominated by
$\tilde{S}_{inc}(\tilde{q},\tilde\omega;\tilde{T})$.

\begin{figure}
\begin{center}
\includegraphics[height=16cm]{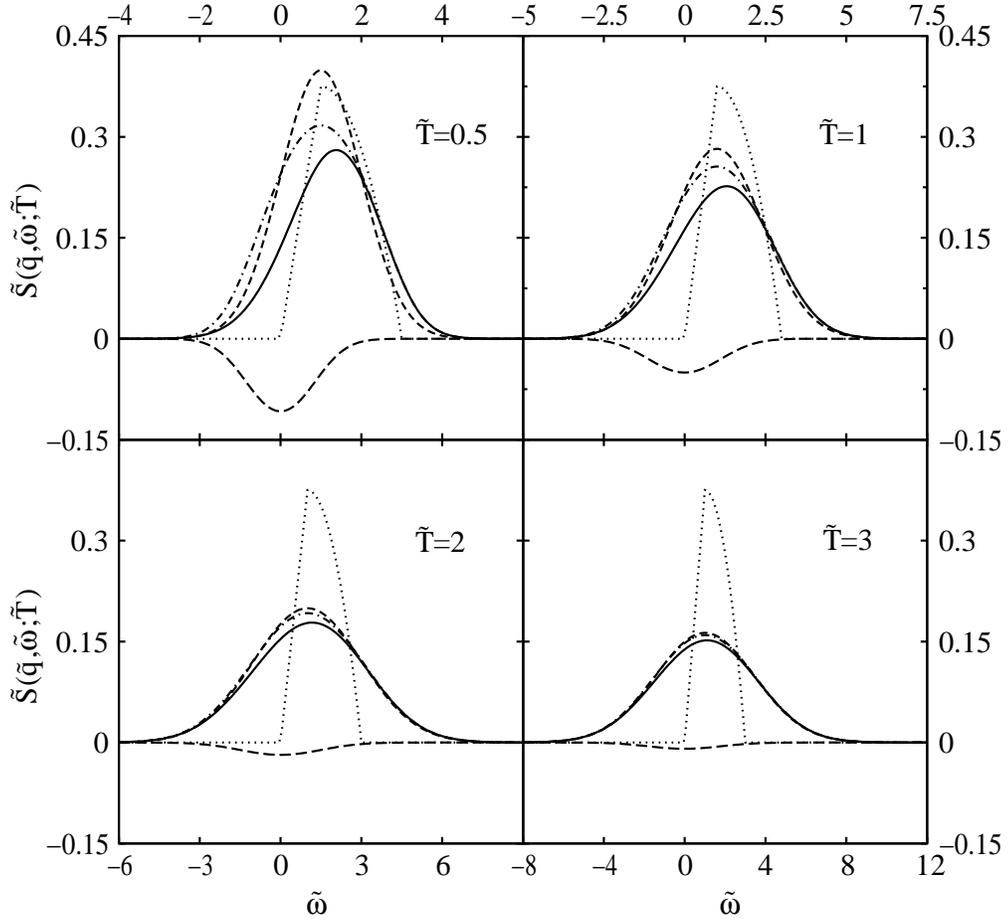}
\caption{Different responses at several temperatures. Solid line:
total response at finite temperature, dotted line: total response at 
$\tilde{T}=0$, dot--dashed line: incoherent response at $\tilde{T}\neq 0$, 
long--dashed line: coherent response at $\tilde{T}\neq 0$, dashed line: 
clasical response at $\tilde{T}\neq 0$.}
\label{fig-3}
\end{center}
\end{figure}

The evolution with $\tilde{T}$ of the total response and its coherent
and incoherent parts is reported in figure~(\ref{fig-3}) for $\tilde
q=1$ and four different values of the adimensional temperature. As at
high $\tilde{T}$ almost all particles have been promoted outside the
Fermi sphere and the mean occupation of every single--particle state
is low, statistical effects are drastically weakened and the system
reaches the classical limit. In this case and due to the absence of
correlations, the response becomes completely incoherent and conforms
to the IA computed from a Maxwellian momentum distribution
\begin{equation}
\tilde S_{cl}(\tilde q, \tilde\omega; \tilde T) = 
{1 \over 2\tilde q \sqrt{\pi \tilde{T}}}\, e^{ -{1\over 4\tilde{T}}
\left({\tilde\omega\over\tilde q}-\tilde q\right)^2} \ ,
\label{sqw-f1}
\end{equation}
a result that can be easily derived from eq.(\ref{sqw-d5}) using the
asymptotic expression of the chemical potential~\cite{pathria}
\begin{equation}
\tilde\mu(\tilde T\!\to\!\infty) \to 
\tilde T \ln\left( \frac{4}{3\pi^{1/2}\tilde{T}^{3/2}}\right) \ .
\label{sqw-f2}
\end{equation}

As it can be seen from the figure, the low temperature coherent,
incoherent and total responses depart from their classical prediction,
i.e.  $\tilde{S}_{coh}(\tilde{q}, \tilde\omega; \tilde T)=0$ and
$\tilde{S}(\tilde q, \tilde\omega; \tilde T) = \tilde{S}_{inc}(\tilde
q, \tilde\omega; \tilde T) \equiv \tilde{S}_{cl}(\tilde q,
\tilde\omega; \tilde T)$.  At finite temperature,
$\tilde{S}_{coh}(\tilde{q},\tilde\omega;\tilde{T})$ is negative and
$\tilde{S}_{inc}(\tilde{q},\tilde\omega;\tilde{T})$ is positive, but
when $\tilde{T}$ is raised the former goes to zero while the latter
approaches the classical limit, a fact that brings the total response
closer to $\tilde{S}_{cl}(\tilde q, \tilde\omega; \tilde T)$ but at a
slower rate.  Notice, however, that the way in which the classical
limit is approached depends on the value of the momentum transfer. At
$\tilde{q}<2$ the contribution of the coherent response is maximal at
$\tilde{T}=0$ and decreases with increasing $\tilde{T}$. At
$\tilde{q}>2$ $\tilde{S}_{coh}(\tilde{q}, \tilde\omega; \tilde{T})$ is
finite at intermediate temperatures but vanishes at $\tilde{T}=0$ and
$\tilde{T}\to\infty$. In this way, temperature brings incoherentness
to the response when $\tilde{q}<2$ and coherentness when
$\tilde{q}>2$. The presence of coherent effects is also reflected in
the shape of the total response, as when $\tilde{S}_{coh}(\tilde{q},
\tilde\omega; \tilde{T})$ is zero $\tilde{S}(\tilde q, \tilde\omega;
\tilde T)$ equals $\tilde{S}_{inc}(\tilde{q}, \tilde\omega; \tilde T)$
and thus the total response becomes maximal and symmetric around
$\tilde\omega=\tilde{q}^2$.  Out of this limit, $\tilde{S}(\tilde{q},
\tilde\omega; \tilde T)$ is asymmetric and peaks at some positive
energy larger than $\tilde{q}^2$, a direct consequence of
$\tilde{S}(\tilde{q}, \tilde\omega; \tilde T)$ being the sum of
$\tilde{S}_{inc}(\tilde{q}, \tilde\omega; \tilde T)$ and
$\tilde{S}_{coh}(\tilde{q}, \tilde\omega; \tilde T)$ while the latter
is negative and minimal at $\tilde\omega=0$. In this sense, both the
asymmetry of the response and the shift of its maximum to slightly
larger energies is a clear signature of the presence of coherent
contributions.


The role played by coherent and incoherent effects at finite
temperature can also be analyzed in terms of sum rules, which are
energy weighted integrals of the different responses
\begin{eqnarray}
\tilde{m}_{inc,coh}^{(\alpha)}(\tilde q; \tilde T) & \equiv & 
\int_{-\infty}^\infty \tilde\omega^\alpha 
\tilde S_{inc,coh}(\tilde q, \tilde\omega; \tilde T)
\,d\tilde\omega
\nonumber \\ [2mm]
\tilde{m}^{(\alpha)}(\tilde q; \tilde T) \equiv 
\tilde{m}_{inc}^{(\alpha)}(\tilde q; \tilde T) & +  &
\tilde{m}_{coh}^{(\alpha)}(\tilde q; \tilde T) = 
\int_{-\infty}^\infty \tilde\omega^\alpha
\tilde S(\tilde q, \tilde\omega; \tilde T) \,
d\tilde\omega \ .
\label{sqw-g1}
\end{eqnarray}

Up to the first two orders, these read
\begin{equation}
\begin{array}{ccc}
\tilde{m}_{inc}^{(0)}(\tilde q; \tilde T) = 1 \,\,\,\,\,& 
\tilde{m}_{coh}^{(0)}(\tilde q; \tilde T) = 
\tilde{S}(\tilde q; \tilde T)-1 \,\,\,\,\, & 
\tilde{m}^{(0)}(\tilde q; \tilde T) = 
\tilde{S}(\tilde q, \tilde T) \\ [2mm]
\tilde{m}_{inc}^{(1)}(\tilde q; \tilde T) = \tilde q^2 & 
\tilde{m}_{coh}^{(1)}(\tilde q; \tilde T) = 0 &
\tilde{m}^{(1)}(\tilde q; \tilde T) = \tilde q^2 \ ,
\end{array}
\end{equation}
where $\tilde{S}(\tilde{q}; \tilde{T})$ is the finite temperature
static structure factor.  The $\tilde{m}^{(1)}(\tilde{q}; \tilde{T})$
sum rules do not yield particular information about the behavior of
the free Fermi gas because they are known to be satisfied by both
correlated and uncorrelated homogeneous and isotropic systems.  On the
other hand, the static structure factor is a positive defined quantity
that in the current case turns out to be smaller or equal to $1$, as
can be inferred from the properties of $n(\tilde{k})$ and the
deffinition of $\tilde{S}(\tilde{q}, \tilde\omega; \tilde{T})$ in
eq.(\ref{sqw-d3b}). The analysis of such an integrated quantity brings
information about how coherent and incoherent effects {\em globally}
affect the total response, in contrast to the previous {\em local}
analysis in which the contribution of $\tilde{S}_{inc}(\tilde{q},
\tilde\omega; \tilde{T})$ and $\tilde{S}_{coh}(\tilde{q},
\tilde\omega; \tilde{T})$ at each $\tilde\omega$ was discussed.

$\tilde{S}(\tilde{q}; \tilde{T})$ can be numerically evaluated and is
shown for several temperatures in Fig.(4). As it can be seen, the
finite temperature and the $\tilde{T}=0$ static structure factors are
different. At low momentum, $\tilde{m}_{coh}^{(0)}(\tilde{q};
\tilde{T})$ is closer to $0$ than $\tilde{m}_{coh}^{(0)}(\tilde{q};
0)$, thus showing that $\tilde{S}(\tilde{q}, \tilde\omega; \tilde{T})$
is globally less coherent that $\tilde{S}(\tilde{q}, \tilde\omega;
0)$. In this way, the visible local reduction in the strength of
$\tilde{S}_{coh}(\tilde{q}, \tilde\omega; \tilde{T})$ at low
$\tilde{q}$'s and $\tilde\omega$'s originates an overall global
reduction of coherentness in the total response (see Fig.(1)).
However, at high $\tilde{q}$ the opposite situation is found. At
$\tilde{T}=0$ the static structure factor is $1$ for all $\tilde{q}>2$
while at finite temperature this value is only asymptotically reached
when $\tilde{q}\to\infty$. In this sense, therefore, and in contrast
to what happens at low momentum transfer, thermal effects induce
coherent contributions at high $\tilde{q}$. Notice, however, that
depending on the value of $\tilde{q}$ this effect may not be easily
appreciated in the plots. This is because at high momentum transfer
the global enhancement of coherentness in the total response is mainly
due to a strong broadening of $\tilde{S}_{coh}(\tilde{q},
\tilde\omega; \tilde{T})$ over the whole energy axis rather than to a
local increase of strength at low energies.


One of the most interesting features of the density response of a
system of interacting particles is the scaling behaviour, i.e. in the
high momentum transfer limit the response becomes mainly incoherent
and is driven by the West scaling variable $\tilde{Y}$. In order to
put in evidence the scaling one introduces the adimensional Compton
profile
\begin{equation}
\tilde{J}(\tilde{q},\tilde{Y}; \tilde{T}) = 2\tilde{q}\,
\tilde{S}(\tilde{q},\tilde\omega; \tilde{T}).
\label{eq:comp1}
\end{equation}

In this context scaling means that $\tilde{J}(\tilde{q},\tilde{Y};
\tilde{T})$ does not depend on $\tilde{q}$. In the free case, the only
source of correlations is the Pauli principle and therefore as stated
before the incoherent response and the impulse approximation
coincide. The adimensional Compton profile, associated to the
incoherent response, at finite $\tilde{T}$ reads
\begin{equation}
\tilde{J}_{IA}(\tilde{Y}; \tilde{T}) = 
{{3}\over{4}}\,\tilde{T} \ln \left(1 + 
e^{-\frac{\left( \tilde{Y}^2-\tilde\mu \right)}{\tilde{T}}} \right)
\label{eq:comp2}
\end{equation}
where variable $q$ has been omitted for being superflous. Notice that
this adimensional expression depends only on $\tilde{Y}$ and
$\tilde{T}$, thus being mass and density independent.
$\tilde{J}_{IA}(\tilde{q}; \tilde{T})$ is also a symmetric function of
$\tilde{Y}$. At $\tilde{T}=0$ this expresion reduces to
\begin{equation}
\tilde{J}_{IA}(\tilde{Y}; 0) = \frac {3}{4}(1 - \tilde{Y}^2) \ ,
\label{eq:compt0}
\end{equation}
defined in the interval $-1 \leq \tilde{Y} \leq 1$. On the other hand,
when $\tilde{T}\to\infty$ one obtains the Compton profile associated
to the classical response
\begin{equation}
\tilde{J}_{cl}(\tilde{Y}; \tilde{T}) = \frac{1}{\sqrt{\pi \tilde{T}}}\,
e^{-\tilde{Y}^2/\tilde{T}} \ .
\end{equation} 

At zero temperature and for $\tilde{q}\geq 2$, the Pauli principle has
no effect and so the total response is completly incoherent and scales
according to Eq.~(\ref{eq:compt0}).  When the temperature incresases,
the incoherent part still scales but due to the Pauli principle the
coherent part of the response is not zero and so the scaling property
of the total response is lost.

Of course, at fixed temperature the scaling regime is recovered when
$\tilde{q}$ increases. This fact is illustrated in Fig.(5) where the
Compton profile associated to the full response
$\tilde{J}(\tilde{q},\tilde{Y}; \tilde{T})$ is reported at $\tilde{T}
=1$ and for different values of $\tilde{q}$. Curves corresponding to
$\tilde{q}=2$ and $\tilde{q}=3$ would coincide at $\tilde{T}=0$, and
thus their difference is a measure of how thermal effects break 
scaling. Already at $\tilde{q}=3$ the asymptotic behaviour has almost
been reached and the Compton profile is very well approximated by
$\tilde{J}_{IA}(\tilde{Y}; \tilde{T})$.

Another interesting point is to study how at fixed $\tilde{q}$ the
response approaches the classical limit $\tilde{J}_{cl}(\tilde{Y};
\tilde{T})$ when temperature increases. However, as
$\tilde{J}_{cl}(\tilde{Y}; \tilde{T})$ itself depends on $\tilde{T}$,
it is useful to eliminate this dependence by introducing a new
variable $\tilde{Y}_{T}=\tilde{Y}/\tilde{T}^{1/2}$ and a new Compton
profile $\tilde{\jmath}(\tilde{q}, \tilde{Y}; \tilde{T})$
\begin{equation}
\tilde{\jmath}(\tilde{q},\tilde{Y}_{T}; \tilde{T}) \equiv
\tilde{T}^{1/2} \tilde{J}(\tilde{q},\tilde{Y}; \tilde{T}) \ .
\label{eq:unicla}
\end{equation}

With this new definition,
\begin{equation}
\tilde{\jmath}_{cl}(\tilde{Y}_{T}) = 
\frac{1}{{\sqrt\pi}}\,e^{-\tilde{Y}_T^2}
\label{eq:jclass}
\end{equation}
which is an universal function valid for all free and classical
systems. In Fig.~(6a) $\tilde{\jmath}(\tilde{q},\tilde{Y}_T;
\tilde{T})$ is plotted at $\tilde{q}=1$ as a function of $\tilde{T}$.
At low temperatures $\tilde{\jmath}(\tilde{q},\tilde{Y}_T;
\tilde{T})$ is remarkably different from 
$\tilde{\jmath}_{cl}(\tilde{Y}_T)$ and not symmetric in
$\tilde{Y}_T$. When $\tilde{T}$ increases, 
$\tilde{\jmath}(\tilde{q},\tilde{Y}_T; \tilde{T})$ 
approaches $\tilde{\jmath}_{cl}(\tilde{Y}_T)$.
In Fig.~(6b) the same comparison is shown for the
incoherent part of the response 
$\tilde{\jmath}_{IA}(\tilde{Y}_T; \tilde{T})$. In this
case the curves are independent of $\tilde{q}$ because 
the latter scales. In addition, 
$\tilde{\jmath}_{IA}(\tilde{Y}_T; \tilde{T})$ is a symmetric
function of $\tilde{Y}_T$ that approaches 
$\tilde{\jmath}_{cl}(\tilde{Y}_T)$ when $\tilde{T}$ increases. 
Notice that these definitions of $\tilde{\jmath}$'s is appropriate only
to study the limit of high temperature. 

In summary, the dynamic structure function of the free Fermi gas has been 
formally separated in its coherent and incoherent parts inorder to analyze
the influence of thermal effects in the total response. The relative 
contribution of the coherent and incoherent responses turns out to 
depend on the momentum transfer. At low $\tilde{q}$ the main effect 
produced by the temperature is to reduce coherentness from the 
response. On the other hand, at high $\tilde{q}$, thermal effects 
act as a source of coherentness. As a consequence, temperature 
breaks the scaling behaviour characteristic of the high $\tilde{q}$ 
response at $\tilde{T}=0$. However, scaling is recovered in the 
$\tilde{T}\to\infty$ limit, where the total response approaches
the classical limit. Despite of the simplicity of the system 
considered, the present analysis can give insight in understanding 
the behaviour of the response of very dilute interacting Fermi 
systems.

\bigskip

This work has been partially supported by grants DGICYT (Spain) 
PB95--1249 and the program SGR98--11 from 
{\em Generalitat de Catalunya}.

\vfill\eject


\vfill\eject


%

\pagebreak


\pagebreak

%

\pagebreak

%

\begin{figure}
\begin{center}
\includegraphics[height=16cm]{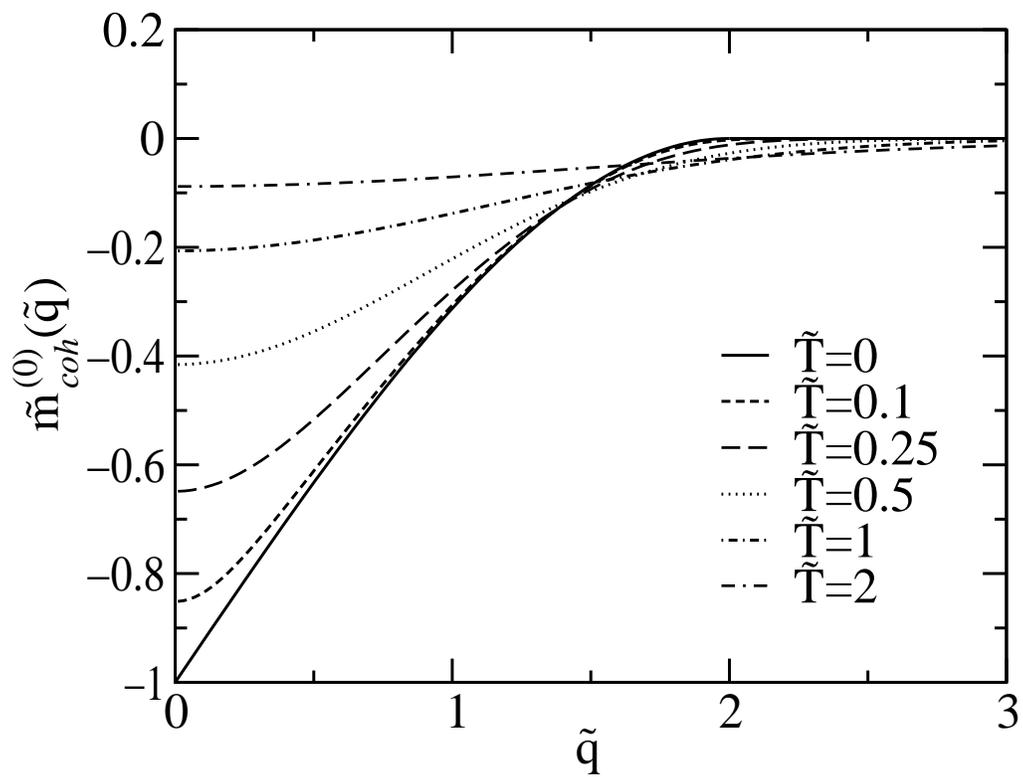}
\caption{Momentum dependence of the zero order sum rule of the coherent
response at different temperatures.}
\label{fig-4}
\end{center}
\end{figure}

\pagebreak

%

\begin{figure}
\begin{center}
\includegraphics[height=16cm]{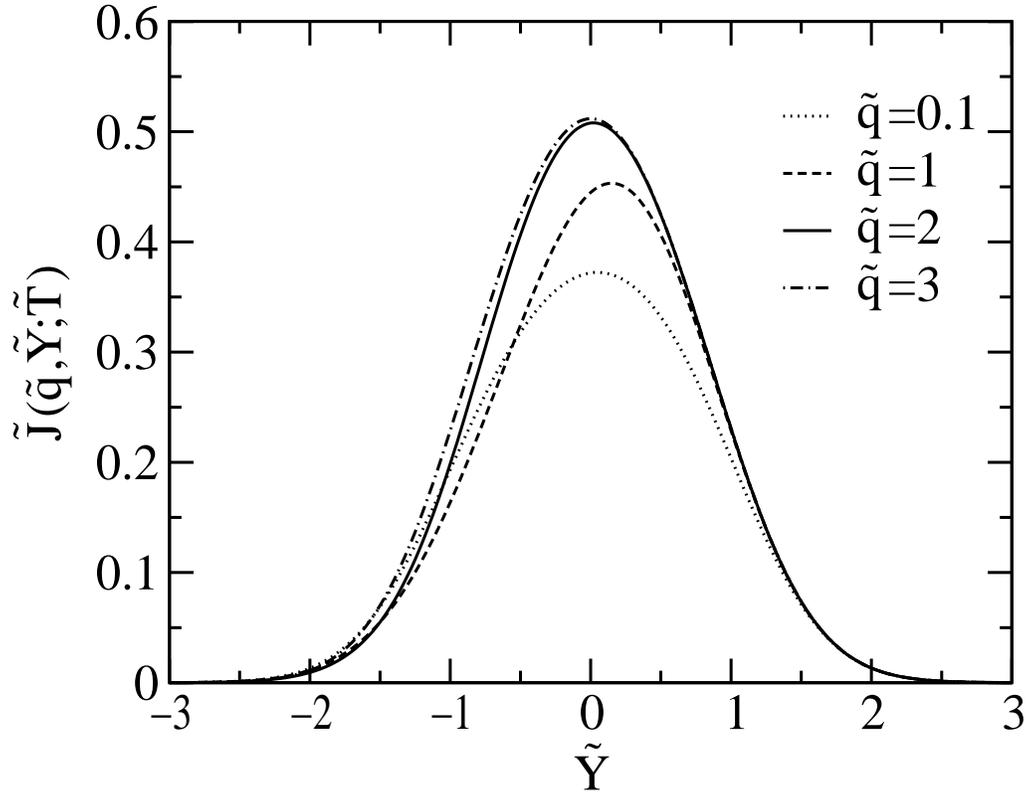}
\caption{Finite temperature Compton profile of the total response
at $\tilde{T}=1$ and several values of the momentum transfer $\tilde{q}$.}
\label{fig-5}
\end{center}
\end{figure}

\pagebreak

%

\begin{figure}
\begin{center}
\includegraphics[height=16cm]{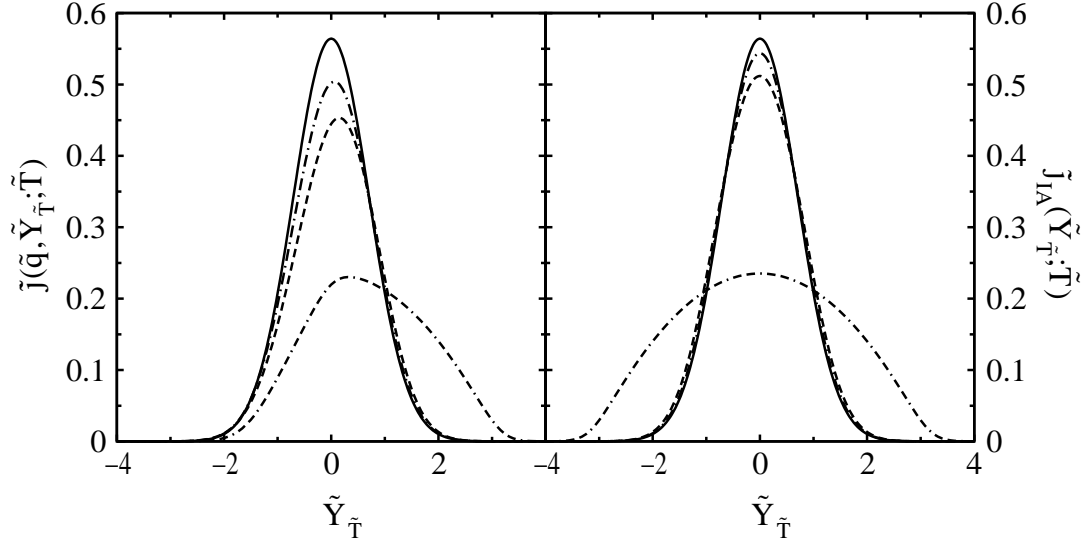}
\caption{(a) $\tilde{\jmath}(\tilde{q}, \tilde{Y}_T; \tilde{T})$ and
$\tilde{\jmath}_{IA}(\tilde{Y}_T; \tilde{T})$ at $\tilde{q}=1$ for
several temperatures. Solid line: universal
$\tilde{\jmath}_{cl}(\tilde{Y}_T)$ (see Eq.~(\ref{eq:jclass}));
dot--dashed line: $\tilde{T}=2$; dashed line: $\tilde{T}=1$, and
dot--dashed--dashed: $\tilde{T}=0.1$.}
\label{fig-6}
\end{center}
\end{figure}

\vfill\eject

\end{document}